# Astro2020 Activities and Projects White Paper: Arecibo Observatory in the Next Decade

July 10, 2019

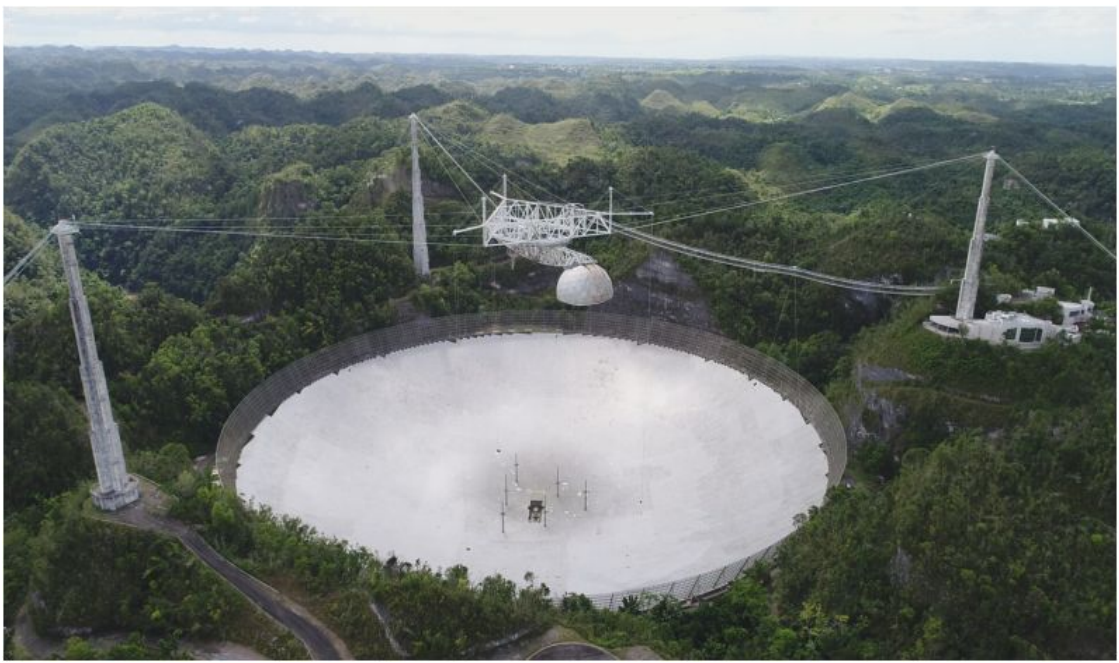

**Contact Author:** D. Anish Roshi,[1] aroshi@naic.edu

**Authors:** L. D. Anderson,[2] E. Araya,[3] D. Balser,[4] W. Brisken,[4] C. Brum,[1] D. Campbell,[5] S. Chatterjee,[5] E. Churchwell,[6] J. Condon,[4] J. Cordes,[5] F. Cordova,[1] Y. Fernandez,[7] J. Gago,[1] T. Ghosh,[8] P. F. Goldsmith,[9] C. Heiles,[10] D. Hickson,[1] B. Jeffs,[11] K. M. Jones,[1] J. Lautenbach,[1] B. M. Lewis,[1] R. S. Lynch,[8] P. K. Manoharan,[12] S. Marshall,[1] R. Minchin,[13] N. T. Palliyaguru,[1] B. B. P. Perera,[1] P. Perillat,[1] N. Pinilla-Alonso,[14,1] D. J. Pisano,[2] L. Quintero,[1] S. Raizada,[1] S. M. Ransom,[4] F. O. Fernandez-Rodriguez,[1] C. J. Salter,[1,8] P. Santos,[1] M. Sulzer,[1] P. A. Taylor,[15] F. C. F. Venditti,[1] A. Venkataraman,[1] A. K. Virkki,[1] A. Wolszczan,[16] M. Womack,[14,7] and L. F. Zambrano-Marin[1]

*Affiliations are listed at the end of the paper*

**Related Science White Papers:**

"Physics Beyond the Standard Model With Pulsar Timing Arrays", Siemens et al. 2019
"Radio Time-Domain Signatures of Magnetar Birth", Law et al. 2019
"Magnetic Fields of Extrasolar Planets: Planetary Interiors and Habitability", Lazio et al. 2019
"Invisible Structures in the Local Universe", Minchin 2019
"Planetary Radar Astronomy with Ground-Based Astrophysical Assets", Taylor et al. 2019
"Radar Astronomy for Planetary Surface Studies", Campbell et al. 2019
"Gravitational Waves, Extreme Astrophysics, and Fundamental Physics with Precision Pulsar Timing", Cordes et al. 2019
"Galactic and Extragalactic Astrochemistry: Heavy-Molecule Precursors to Life?", Heiles et al. 2019
"Technosignatures in Transit", Wright & Kipping 2019
"Magnetic Fields and Polarization in the Diffuse Interstellar Medium", Clark et al. 2019

# 1.0 Introduction

The Arecibo Observatory (AO) has supported cutting-edge research in the fields of Astronomy, Planetary Science, and Space Atmospheric Science. The key strengths of the Arecibo Telescope are the highest attainable instantaneous sensitivity up to 10 GHz from its large collecting area (305 m diameter) and its sophisticated and flexible data acquisition and signal processing systems (Altschuler 2002; Altschuler & Salter 2013). Arecibo Telescope is still the largest and most sensitive single dish telescope in the world at frequencies between 2 and 10 GHz. AT's unprecedented sensitivity has led to fundamental contributions in a wide variety of research programs, including the first detection of an exoplanet around pulsar (Wolszczan & Frail 1992), indirect detection of gravitational waves (GWs) that resulted in the 1993 Nobel Prize in Physics (Hulse 1994), and discovery and localization of a repeating Fast Radio Burst (Spitler et al. 2016, Marcote et al. 2016).

Major upgrades of the telescope took place in the 1990s with the installation of the broad-band Gregorian system and a high powered transmitter for planetary radar studies (Goldsmith 1996). More recent instrumentation enhancements include the installation of the 7-horn Arecibo L-band Feed Array (ALFA) system in 2004 (Heiles 2004) and ionospheric heating facility commissioned in 2015 (Breakall 2013). The next generation of upgrades to the Arecibo telescope are critical to keep this national facility in the forefront of research in radio astronomy while maintaining its dominance in radar studies of near-Earth asteroids, planets and satellites.

In February 2019, the AO organized a 3-day workshop titled "Pathways to the Future of the AO" in San Juan, Puerto Rico[1]. The purpose of the workshop was to create a shared vision of the future research that can be conducted using the telescope and to discuss the potential technical upgrades that would ensure that the AO stays at the forefront of radio astronomy, space and atmospheric sciences, and planetary sciences research. The community had numerous suggestions for improvements of the AT's capabilities; many of these suggestions concur with the short term priorities and equipment development plans proposed by the Arecibo Users Committee[2]. We broadly categorize the suggestions into (a) 'facility improvements', which will be implemented in the next decade (see Fig. 1) and (b) 'major upgrades' that require feasibility studies and community science build up before implementation. We briefly discuss the facility improvements in Section 2 and plans for preparation for the major upgrades in Section 4. A few other important activities during the next decade are summarized in Section 3.

# 2.0 Facility improvements

### 2.1 Advanced L-band Phased Array Camera for Arecibo (ALPACA)

The ALPACA will be a 40-beam cryogenic Phased Array Feed (PAF) facility instrument for the Arecibo Telescope operating over the frequency range 1300 to 1720 MHz. The project is fully funded by the National Science Foundation and is being developed by Brigham Young University (BYU) and Cornell University in partnership with the Arecibo Observatory. The beams are formed by sampling the focal field using 69 dual-polarized dipole elements and then processing these antenna signals with a real-time digital array beamformer backend. Beam spacing will be close to Nyquist separation on the sky to provide continuous coverage of the field-of-view of ~340 square arcmin. Unique features of ALPACA, which will provide significant performance increases for a number of important science cases are: a) an unmatched combination of sensitivity and wide field of view for a single dish instrument; b) the first fully cryogenically cooled (both antennas and LNAs) L-band radio astronomical PAF; c) continuous field of view coverage on this class of telescope by digital beamforming; and d) potential for future development of active spatial filtering to cancel Radio Frequency Interference (RFI) with adaptive beam nulling (Jeffs et al. 2019). ALPACA development work was started in mid 2018 and installation on the telescope is scheduled for 2022.

---

[1] http://www.areciboobservatory.org/futures/
[2] Arecibo Users Committee Report, 2018

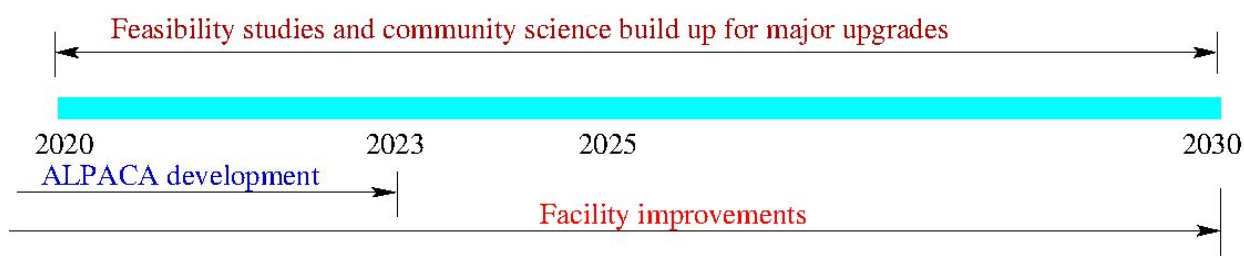

Figure 1: Timeline for facility improvements and feasibility studies for major upgrades over the next 10 years.

**Key science goals:** ALPACA will survey the neutral hydrogen (HI) content of nearby groups and clusters, giving an unprecedented combination of column density and mass sensitivity. Surveys with ALPACA will build on the previous generation of HI surveys, performed at AO with ALFA, but will obtain statistically significant samples of interesting objects that had only a few detections in the earlier surveys. Targets will include low mass galaxies (Giovanelli et al. 2010), low metallicity galaxies similar to Leo P (Skillman et al. 2013), and optically-dark HI clouds similar to those seen in Virgo (Minchin et al. 2019). The opportunity to tune the beam shape of ALPACA enables high fidelity imaging of extremely low HI column densities, $< 10^{17}$ $cm^{-2}$. This imaging capability permits detection and mapping of faint HI structures around nearby galaxies (e.g. Pisano 2014). Such features will reveal past tidal interactions as well as ongoing accretion from the circumgalactic medium (e.g. Keres et al. 2005, 2009). The ALPACA will also have a major impact on Galactic HI absorption/emission and Zeeman measurements (Clark et al. 2019).

ALPACA will provide advanced survey speed for pulsars, and the same data can be used to detect fast radio bursts (FRBs). The most important goals for pulsar surveys are to (1) find high-quality millisecond pulsars (MSPs) for inclusion in the NANOGrav pulsar timing array detector of nanohertz gravitational waves; (2) find relativistic binary pulsars with neutron star and potentially black-hole companions that will provide precision masses and tests of general relativity; and (3) provide additional lines of sight for characterizing the magneto-ionic interstellar medium. ALPACA discoveries will lead to follow-up observations to determine whether they have binary companions and what their overall timing quality is. While ALPACA's discovery rate for FRBs will not be as high as that of wide-field instruments (e.g. CHIME Telescope, Canada), ALPACA will play an important role in localizing repeating FRB sources that have poor localizations and for probing spectro-temporal structure in any repeating bursts.

## 2.2 Improving the telescope surface, pointing, and focusing

The performance of the Arecibo Telescope at frequencies above ~2 GHz depends critically on a) the degree to which the primary reflector surface conforms to its designed geometry and b) the accuracy of the pointing and focusing while tracking the target source. In this section we discuss our strategies to improve the system performance at higher frequencies (up to 12.5 GHz) and key science opportunities this will present.

**Key science goals:** High sensitivity observations at cm wavelengths can result in unambiguous detections of heavy molecules that are precursors to life in the ISM, including polycyclic aromatic hydrocarbons, which may be practically impossible to identify at high frequencies due to line-forests at mm/sub-mm wavelengths (Heiles et al. 2019). The detections of carbon-chain molecules such as $C_4H$ at 9.5 GHz toward TMC-1 (Kalenskii et al. 2004) exemplifies the potential of the telescope for such studies. Moreover, the detection of methanimine toward Arp 220 (Salter et al. 2008) demonstrates the depth of discovery space of the telescope in the area of extragalactic astro-chemistry. Pulsar observations at relatively high frequencies, especially for distant pulsars (as we expect most discoveries to be), may benefit from minimized variable scattering and time delays due to the turbulent ionized interstellar medium. For example, Arecibo pulsar J1903+0327 would be better observed at a much higher frequency than is presently done (Lam et al. 2018).

A combination of high telescope gain at frequencies above a few GHz, wide-band receivers, backend capabilities, and pointing accuracy will allow sensitive studies of several spectral transitions which include: a) higher excitation transitions of OH (e.g. $2\Pi_{1/2}$ J=3/2 lines at 7.8 GHz, $2\Pi_{1/2}$ J=5/2 lines at 8.1 GHz), for example, from expanding gas in pre-planetary nebula (Strack et al., 2019); b) K-doublet transitions of $H_2CO$ at 6 cm for densitometry research (e.g., Ginsburg et al. 2015); and c) masers from multiple molecular species (OH, $H_2CO$, 6.7 and 12.2 GHz transition of $CH_3OH$; e.g., Al-Marzouk et al. 2012) that exhibit periodic flares (e.g., Goedhart et al. 2007; Szymczak et al. 2018), which may trace periodic accretion events in binary systems, proto-stellar pulsations, periodic enhancements of ionized wind shocks in young eccentric binaries, among other possibilities (e.g., Araya et al. 2010).

Radar measurements of Near-Earth Asteroids (NEAs) are fundamental to support planetary defense and our knowledge of small bodies in the Solar System. The AO planetary radar system operates in the S-band (2.38 GHz), and is the most powerful and sensitive radar system in the world (Naidu et al. 2016), observing roughly 100 NEAs per year. As such, AO is vital for post-discovery physical and dynamical characterization of NEAs and for accomplishing the federal mandates of tracking and characterization set forth by the *George E. Brown, Jr. Near-Earth Object Survey Act* and the *National Near-Earth Object Preparedness Strategy and Action Plan* (Taylor et al. 2019). Any improvement in the telescope gain at this frequency will have a positive impact on the radar system performance as it depends on the square of the aperture efficiency.

**Instrumentation:** The primary reflector for the Arecibo Telescope is a spherical surface composed of 38,788 perforated aluminium panels with fabrication RMS error of ~0.1 cm. The panels, which are supported by a grid of cables, are individually adjustable and the surface shape is maintained by `tieback' cables attached to the main grid of cables. The surface is adjusted so that it follows a spherical geometry. The surface deforms from an ideal spherical surface due to a variety of reasons, which include (a) errors in panel setting and (b) deviations due to soil motions at the base of the telescope. The panels have been last reset using photogrammetric measurement in 2001. The measured primary surface RMS error after adjusting the panels was ~0.185 cm and the aggregate surface RMS error is ~0.25 cm. This aggregate error is consistent with the measured telescope gain. The error contribution from the secondary reflector is ~0.12 cm (see Goldsmith 2002 for further details).

The surface deformation caused after Hurricane Maria in 2017 has significantly reduced the telescope gain. AO is currently investigating techniques which include laser scanning and photogrammetry to make fast (within an hour) surveys of the surface to an accuracy of 0.1 cm. The current priority is to permanently install such a measurement system on the telescope, make the surface survey, and correct for the deformation. Further, earlier analysis of the surface error contributions indicates that the aggregate error could be reduced by (a) improving the secondary panel setting and (b) fine adjustment of the primary panels (Goldsmith 2002). High-speed photogrammetry can be used to accurately measure the secondary surface, as it is small (~ 20 m) and located in the dome, which will help to adjust its panels to reduce the surface errors. Repeated primary surface surveys and iterative adjustment of the panels, which will be now possible with fast measurement systems, will help in reducing the primary surface errors. Such repeated measurements will also be useful to a) measure the diurnal and temperature dependent variations of the panel setting and b) possibly measure the RMS error of individual panels and replace the deformed panels. The improvement in telescope gain for different aggregate RMS values is shown in Fig. 2. If the net RMS error can be reduced to 0.19 cm (error contributions: primary panel 0.1 cm, primary surface 0.15 cm, secondary panel 0.025 cm and secondary surface 0.05 cm) then the expected improvement in telescope gain at 10 GHz is about 50% and the gain at 12 GHz will be 4.5 K/Jy.

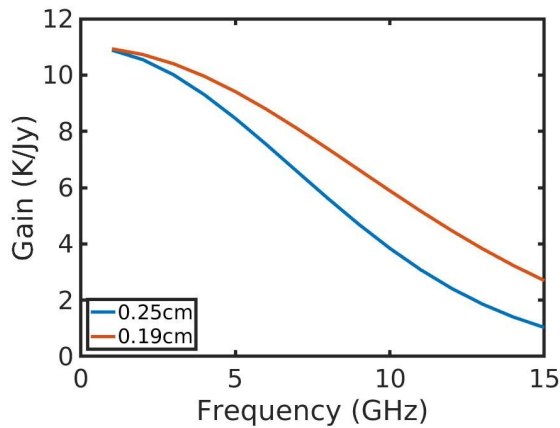

Figure 2. Telescope gain as a function of frequency for different values of the aggregate RMS surface error. The RMS error after the 2001 panel adjustment is about 0.25cm. During the facility

improvements phase we will attempt to reduce the surface errors to 0.19 cm (see Sec 2.2), which would provide a telescope gain of 4.5 K/Jy at 12 GHz.

The pointing and focusing of the telescope need to be improved along with surface correction for high frequency observations. The receivers are located in the Gregorian dome. The dome is mounted on the azimuth arm which in turn is suspended from a large supporting platform. The total weight of the dome is ~100 tons, which introduces a large asymmetry in the distribution of weight on the platform. As a result of this weight asymmetry the platform translates and tilts as the dome moves while tracking an astronomical source. This motion of the platform introduces pointing and focusing errors, which need to be corrected for high frequency operations. In the current system a 'tiedown' mechanism is used along with feedback from a laser ranging system (distomats) and 'tilt sensors' to correct for the platform motion. AO plans to upgrade the telescope instrumentation (installing modern azimuth and zenith control system, distomats, laser scanner, tilt sensors, and tiedown controller) to accurately measure the pitch, roll, and tilt of the platform and correct for its motion. Further, we plan to install triaxial movement capability to the tertiary reflector, which will help in precise focusing and to some extent compensate for the sag of the platform. The data from the upgraded instrumentation will also help in developing a more sophisticated pointing and focusing model.

## 2.3 Wide-band receivers

AO plans to install/upgrade a suite of wide-band receivers which will provide continuous frequency coverage from 0.7 to 12 GHz. This instrumentation will keep the telescope internationally competitive and facilitate new science opportunities. Replacing existing receivers with wideband systems will also reduce the number of receivers, thus reducing the maintenance cost. Fig. 3a summarizes the planned feed configuration on the turret.

**2.3.1 Ultra-wideband feed (0.7 to 4 GHz):** In the next decade, nanohertz GWs from supermassive black hole binaries will be detected using pulsar timing arrays (PTAs) (Cordes et al. 2019). This technique uses the clock-like nature of pulsars to detect deviations in the arrival time of pulses caused by the influence of GWs that are on the order of tens of nanoseconds over a span of many years. One of the largest sources of systematic error for PTAs is time-variable delays in pulse arrival time caused by fluctuations in the ionized interstellar medium. These delays must be measured with an accuracy of one part in $10^5$ at each epoch using observations at widely separated radio frequencies. A sensitive ultra-wideband (UWB) receiver covering a few hundred MHz to a few GHz will enable instantaneous, high precision measurements of dispersive delay with high observing efficiency. This precision measurement, combined with the higher signal-to-noise obtained through the increased bandwidth, has the potential to double PTA timing precision and increase sensitivity to GWs.

An UWB system would benefit other experiments such as a) high time-precision measurements testing fundamental physics, including the theory of General Relativity (Kramer et al. 2006); b) wide-band spectro-temporal observations of repeating FRBs and other fast transients; c) characterizing the spectra of solar bursts (e.g., type-II and type-III bursts), solar transients, and the frequency structure of interplanetary scintillation; and d) detecting weak Zeeman signature of radio recombination lines from H II regions (ionized gas of atomic hydrogen), and photodissociation regions by averaging a large number of transitions that can be observed over the wide-band.

**Instrumentation:** Typical feeds designed for radio astronomy applications have less than an octave of bandwidth. New design techniques are required to extend the feed bandwidth to more than 1:5 in frequency ratio, which is required for the UWB science application. A quad-ridged horn design with corrugated skirt and a central dielectric spear was demonstrated to have performance similar to a narrow-band feed over the frequency range 0.7 to 4 GHz (Dunning et al. 2015). The corrugated skirt and the dielectric spear improve the beam performance at the lower and higher frequencies respectively. Arecibo plans to procure such an UWB feed and the associated analog electronics. The analog receiver will be designed to have large dynamic range for it to operate linearly in the RFI environment at the telescope site. The output of the receiver will be digitized using 12-bit ADC near the front-end and the data will be transported to the backend through a digital fiber link (see Section 2.4).

**2.3.2 Wide-band feed (4 to 8 GHz):** Radio detections of ultracool dwarfs (UCD) provide insight into their magnetic fields and the dynamos that maintain them. Radio flares from UCDs are circularly polarized and observations indicate that the magnetic field strengths are in kiloGauss range. The flares from such high magnetic field strength systems are detectable at frequencies ~5 GHz. The superb sensitivity of the telescope and the availability of wide-band, linearly polarized receivers are ideal for conducting surveys of flaring radio emission from UCDs. (Further spectroscopic science cases for a wide-band receiver near 5 GHz are included in Section 2.2)

**Instrumentation:** The AO has acquired a wideband feed operating over the frequency range 4 to 8 GHz which is being commissioned. To fully commission the system requires a 4 GHz bandwidth IF system. The IF system for this feed will be designed to make it compatible with the universal 4 GHz bandwidth digital link we plan to install (see Section 2.4). Further, the feed will be upgraded with a polarization switch that will provide native linear or circular polarization for observations. The polarization selection is required for the key science projects and to make it compatible for VLBI applications.

**2.3.3 Wide-band 4-12 GHz Receiver:** AO plans to extend the frequency of operation up to 12.5 GHz to support the science cases described in Section 2.2. Currently, the highest frequency receiver of the telescope can operate over the frequency range of 8 to 10 GHz (X-band feed). The plan is to upgrade the X-band receiver with a 4-12 GHz system. This project will be undertaken after the successful demonstration of the improvements in telescope gain, pointing and focusing at the highest frequencies.

## 2.4 High dynamic range universal data link and backend

**Key science goals:** Achieving the science goals for the wide-band feeds described in Section 2.3 requires a high dynamic range digital link and backend that can transport and process bandwidth of at least 4 GHz. The backend will have the signal processing power to provide commensal observing capability for the ALPACA project. This capability will allow commensal surveys for pulsars, FRBs, other kinds of transient signals, technosignatures, and recombination line surveys.

This commensality can provide thousands of hours of telescope time that otherwise would not be available if these targets had to be surveyed individually.

**Instrumentation:** The signals from the front-end are currently brought to the existing backend through RF over fiber. The worsening RFI situation at the AO demands a high dynamic range digital link which digitizes the signals with 12-bit (or more) ADCs located close to the front-end and transports the data through a fiber optic link to the backend. Such a link would also help to improve the spectral baseline stability. To achieve the science goals of the wide-band feeds, the bandwidth of this link needs to be at least 4 GHz. Also this link will be designed to share data transfer from all receivers. Moreover, the system will have the capability to transport signals from reference antennas for RFI mitigation work.

AO plans to upgrade the data processing system with a 'universal backend' that can process up to 4 GHz bandwidth. This backend will have the capability to support spectroscopic observations (spectral resolution < 0.5 km/s), coherent and incoherent real-time pulsar data processing, technosignature search, and VLBI data recording as well as planetary and atmospheric radar observations. A disk array attached to the backend will form the local data storing system. The data will then be transferred to a 'cloud-based storage' for archival. A simplified block diagram of the proposed system is shown in Fig. 3b.

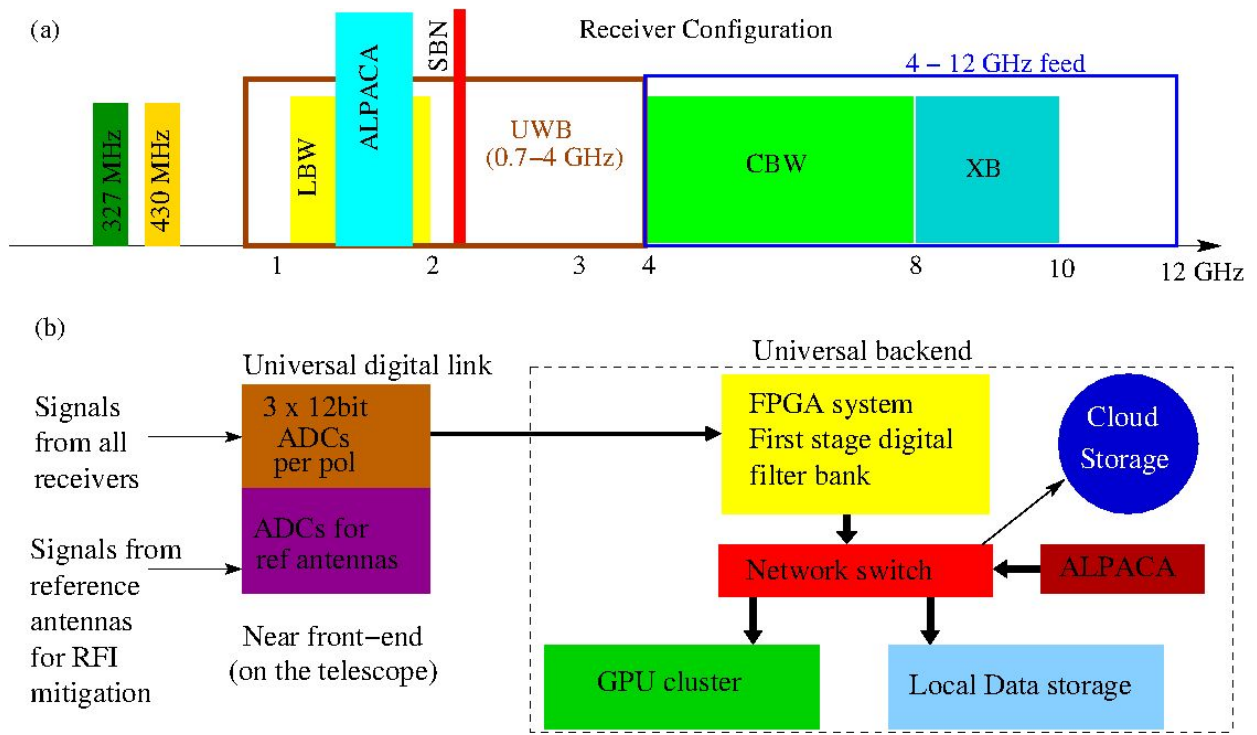

**Figure 3. (a)** Receiver configuration on the turret after the facility improvements (x-axis is not to scale). The receiver frequency ranges in GHz are LBW: 1.15-1.73, SBN: 2.33-2.43, XB: 7.8-10.2, UWB: 0.7-4, ALPACA: 1.3-1.72, CBW: 4-8 and 4-12 feed. **(b)** Block diagram of the 4 GHz (total bandwidth) high dynamic range Universal digital link and Universal backend, which includes a local storage and Cloud-based archival option. The ALPACA beamformer output will be connected to the Universal

backend, which enables commensal observing capability with ALPACA. The block diagram shown here is adopted from a similar system commissioned for Parkes telescope by CSIRO, Australia (G. Hobbs 2019, personal communication).

## 2.5 Very Long Baseline Interferometry (VLBI)

Arecibo Telescope is one of the key components of both the US-based High-Sensitivity Array (HSA) and the European VLBI Network (EVN); it is also critical for any future space-based VLBI. AO is currently upgrading the VLBI system to Mark6 with the two Roach Based Digital Electronics (RDBE) personalities: Polyphase Filter Bank (PFB) and Direct Digital Converter (DDC8). This upgrade will provide a total bandwidth of up to 2 GHz with an aggregate maximum data rate of 4 Gsps. System software will be upgraded to include the option to 'e-ship' the data to the EVN correlator station. Disk packages will be shipped to Socorro for HSA data processing. In the next upgrade step, we plan to use the universal backend (see Section 2.4) for the VLBI data recording.

**12-m telescope:** AO acquired a 12-m antenna to enhance the VLBI capabilities of the Arecibo telescope. This will eliminate the continual need to slew the telescope between a target and calibration source during phase-referenced observations. AO plans to equip the 12-m antenna with cryogenic receivers and integrate it with the VLBI system. The 12-m dish will be able to also operate independently either as single dish telescope or part of VLBI network.

The main science drivers for the 12-m operation include: a) astrometry of weak pulsars and b) high sensitivity (~1microJy/beam) detection of FRBs, extragalactic supernovae remnants, and gamma ray bursts. Phase referencing is essential for both these observations. Other applications are participation in international VLBI service for geodesy and astrometry and, as a stand alone telescope, search for FRBs.

## 2.6 RFI monitoring, control, and mitigation

Radio observatories must control the incidence of man-made RFI (both self-generated and due to other services), and maintain robust procedures for handling RFI in their environs. These tasks are even more crucial now that broad-band receivers cover several GHz of spectrum. Moreover, every observatory has an RFI environment particular to its location that can evolve in time. Consequently the first priority of RFI mitigation is to engineer robust receivers capable of delivering a linear response to received power; the second to provide sufficient headroom to enable the potential for real-time cancellation of RFI in the backend. The high-dynamic range universal digital link and backend (see Section 2.4) are steps that will be taken to implement robust receivers capable of implementing RFI mitigation algorithms. In addition, AO is collaborating with external agencies to support the development of such mitigation algorithms.

Although the AO is located in the Puerto Rico radio astronomy coordination zone, observations are never fully protected from RFI. Routine monitoring of RFI and maintenance of legal limits on emissions are essential for keeping operations free from RFI. AO has long participated in spectrum management at the US national level, and in Geneva at the International Telegraphic Union (ITU). This activity seeks to ensure that spectral bands allocated to the Radio Astronomy service are as free as possible of RFI generated by other spectrum users, via international treaties negotiated at the ITU. These bands are needed to check the validity of broad band continuum observations, as well as to enable spectral line observations, such as redshifted 21 cm line of HI and OH lines.

### 2.7 The search for technosignatures

The search for and understanding of life beyond Earth has a renewed context in the current era of astronomy that has seen significant advances in our understanding of stellar evolution, exoplanets, protoplanetary disks, and planet formation. Thus the search for "technosignatures" of such life can be considered complementary to these other fields (Wright and Kipping 2019). Several of the AO facility improvements discussed above have relevance to technosignature searches: the ALPACA will boost sky coverage by a large factor at L-band; improvements of telescope surface significantly increase the sensitivity; wide-band receivers allow for more spectral territory to be searched; the universal backend will handle baseband data over the whole bandwidth and support the search for a range of signal types; and enhanced data storage will provide capability for future data-mining work.

## 3.0 Other AO activities over the next decade

**Synergy with ngVLA, SKA and FAST:** The two major radio facilities on the horizon are the next generation Very Large Array (ngVLA; https://ngvla.nrao.edu/) and the Square Kilometer Array (SKA; https://www.skatelescope.org/). Arecibo Telescope will form one of the stations for high-sensitivity VLBI observations with ngVLA and SKA. Further, current plans for the ngVLA Long Baseline Array include 3 antennas to be hosted at or near the AO site. The operating frequency range of the recently commissioned Five-hundred-meter Aperture Spherical radio Telescope (FAST) is limited to ~3 GHz. AT's wider frequency coverage is crucial to provide high sensitivity, complementary data at frequencies above 3 GHz. Such complementary observations will benefit, for example, multi-frequency technosignature search, pulsar studies, and pre-biotic molecular line search.

**Broader Impact:** Although interferometers represent the future of radio astronomy, sensitive single-dish telescopes are necessary to identify sources of interest and to fill the missing short-spacing visibilities. Furthermore, single-dish telescopes provide valuable hands-on experience for student researchers, which is not possible for arrays where the reduced data are given to observers. AO will continue supporting student mentoring and training programs such as NSF-REU and RET opportunities and the Single Dish workshop. In addition, AO will grow its support of NANOGrav by supporting the NANOGrav Student Teams of Astrophysics Researchers (NANOStars), Arecibo Remote Command Center (ARCC) training program, and training related to the Arecibo Legacy Fast ALFA survey (ALFALFA) and Arecibo Pisces-Perseus Supercluster Survey (APPSS). These programs reach hundreds of students from diverse backgrounds annually, providing training in observations and data analysis. Funding for these training programs and continued access to open skies time will ensure the growth of the next generation of astronomers and broaden the participation in scientific discoveries.

## 4.0 Preparation for major telescope upgrades

At the 2019 AO Futures workshop, the community discussed possible major upgrades for the telescope. In the next decade, AO plans to undertake a feasibility study of three suggested upgrades: a) extending the low frequency limit from 327 to 30 MHz; b) increasing the frequency of operation above ~12 GHz and c) increasing the declination coverage. These upgrades were selected based on the scientific requirements discussed in different Astro2020 white papers. A number of instrumentation challenges has to be overcome for implementing each of these upgrades, which include: a) feed and robust receiver design and RFI mitigation for the low-frequency system and b) major redesign of the telescope surface, pointing, dynamic focusing, tracking systems and extending the azimuth arm in a cost effective way to increase the frequency coverage above 12 GHz and extend the declination range. Along with the feasibility studies, AO will build up a science community to participate and contribute to these studies and who will eventually become active users of the upgraded Arecibo telescope.

## 5.0 Cost Estimate

The ALPACA (see Section 2.1) is a fully funded project. Most of the cost for instrumentation to improve the surface RMS error, pointing, and focusing discussed in Section 2.2 is accounted in the Hurricane Relief Fund provided for the AO. The cost estimates available for the remaining instrumentation projects as part of the facility improvements (see Section 2) are listed in Table 1. Based on the knowledge of prior projects of similar scope and observatory engineering experience, the total cost required for facility improvements is $6.4 M and falls in the category "small'' provided by the white paper guideline. For the successful completion of facility improvements a strong team of scientific staff and an additional 8 engineering and 2 technical staff need to be supported at AO over the next decade. If the feasibility studies of major upgrades discussed in Section 4.0 lead to a viable, cost effective technical solution, then AO would require major funding estimated as "small'' for upgrade (a) and "medium'' for (b) and (c ) (see Section 4.0).

Table 1: Cost Estimate: Facility Improvements (see Section 2)

| No. | Item | Cost | Category |
|---|---|---|---|
| 1 | Instrumentation for Tertiary motion (see Section 2.2) | 100K | |
| 2 | UWB feed + high dynamic range analog receiver (see Section 2.3.1)[a] | 1.5M | |
| 3 | 4-8 GHz feed improvement + its IF system (see Section 2.3.2) | 300K | Small ($6.4M) |
| 4 | 4-12 GHz feed + its IF system (see Section 2.3.3) | 1.8M | |
| 5 | Universal digital link + backend (see Section 2.4)[a] | 1.5M | |
| 6 | 12m telescope cryogenic receivers (see Section 2.5) | 700K | |
| 7 | RFI monitoring + mitigation work (see Section 2.6) | 450K | |

[a] Cost estimate provided by CSIRO, Australia, based on a similar system commissioned at Parkes Observatory

**Acknowledgement:** We acknowledge the support of the NSF for the "Pathways to the Future of the Arecibo Observatory" workshop through an award from the NSF/ AGS - Upper Atmospheric Facilities Call, 2019. We thank Dick Manchester, David Nice, and Rebecca Koopmann for providing very helpful comments during the preparation of the white paper. The Arecibo Observatory is a facility of the National Science Foundation operated under a cooperative agreement by the University of Central Florida (UCF), Universidad Ana G. Méndez - Recinto de Cupey (UAGM-Cupey), and Yang Enterprises Inc.

The AO Management team (AOMT), led by the UCF, has been managing the facility since April 2018. The team has identified different strategic areas that need to grow, including increasing the scientific staff. As a result, the radio astronomy group has expanded from two to five staff members by July 2019, and continues to grow to strengthen the group's expertise especially in instrumentation, pulsar research, and VLBI. There is also an effort from AOMT to recruit highly skilled post-doctoral researchers through the Pre-eminent Postdoctoral Program at UCF in the future, for example, to work on the 12-m antenna. The head of the radio astronomy group will lead, together with the Science manager (hiring in process), the efforts to stay in contact with the science community of users to optimize the scientific service that AO can offer.

## 6.0 References

Al-Marzouk A. A. et al., 2012, ApJ, 750, 170
Altschuler D. R., Salter C. J., 2013, Physics Today, 66, 11, 43
Altschuler D. R., 2002, "The National Astronomy and Ionosphere Center's (NAIC) Arecibo Observatory in Puerto Rico". Single-Dish Radio Astronomy: Techniques and Applications, ASP Conference Proceedings, 278. Ed. Snezana Stanimirovic, Daniel Altschuler, Paul Goldsmith, and Chris Salter. Astronomical Society of the Pacific, 1
Araya E. D., et al., 2010, ApJL, 717, 133
Breakall J. K., 2013, http://www.naic.edu/~astro/ao50/Arecibo_50th_Paper_Breakall_revised_Oct_23_2013.pdf
Campbell B., et al., 2019 BAAS, 51, 350
Cordes J. M., et al., 2019, BAAS, 51, 447
Dunning A., et al., 2015, in 2015 IEEE-APS Topical Conference on Antennas and Propagation in Wireless Communications (APWC). pp 787–790, doi:10.1109/APWC.2015.7300180
Ginsburg A., et al., 2015, A&A, 573, A106
Giovanelli R., et al., 2010, ApJ, 708, L22
Goedhart S., et al., 2007, Astrophysical Masers and their Environments, 242, 97
Goldsmith P., 2002, http://www.naic.edu/~astro/aotms/performance/AO_perf_oct02.shtml
Goldsmith P. F., 1996, The second Arecibo upgrade, IEEE Potentials, 15(3), 38
Heiles C., 2004, Accurate Parametric Representation of ALFA Main Beams and First Sidelobes: 1344 MHz–1444 MHz, Tech. Rep., NAIC
Heiles C., et al., 2019, BAAS, 51, 376
Hulse R. A., 1994, Review of Modern Physics, 66, 3, 699

Jeffs B.D., et al., 2019, "The Advanced L-Band Phased Array Camera for Arecibo (ALPACA): Design, Capabilities, and Status" Arecibo Observatory Futures Workshop, San Juan, Puerto Rico, Feb. 17-20, 2019. Invited presentation
Kalenskii S. V., et al., 2004, ApJ, 610, 329
Keres D., et al., 2005, MNRAS, 363, 2
Keres D., et al., 2009, MNRAS, 395, 160
Kramer M,. et al., 2006, Science, 314, 97
Lam M. T., et al., 2018, ApJ, 861, 12
Law, C., et al. 2019, BAAS, 51, 319
Lazio, J., et al. 2019, BAAS, 51, 135
Marcote B., et al., 2017, ApJ, 834, L8
Minchin R. F., 2019, BAAS, 51, 524
Minchin R. F., et al., 2019, AJ, submitted
Naidu S. P., et al., 2016, AJ, 152, 99.
Pisano D. J., 2014, AJ, 147, 48
Taylor, P. A., et al., 2019, BAAS 51, 539,
Salter C. J., et al., 2008, AJ, 136, 389
Siemens, X., Hazboun, J., Baker, P.~T., et al. 2019, BAAS, 51, 437
Skillman E. D., et al., 2013, AJ, 146, 3
Spitler L. G., et al., 2016, Nature, 531, 202
Strack A., et al., 2019, ApJ, 878, 90
Szymczak M., et al., 2018, MNRAS, 474, 219
Wolszczan A., Frail D. A., 1992, Nature, 355, 145
Wright J. and Kipping D., 2019, BAAS, 51,343

[1]Arecibo Observatory, HC3 Box 53995, Arecibo, PR 00612, USA
[2]Department of Physics and Astronomy, West Virginia University, P.O. Box 6315, Morgantown, WV 26506, USA
[3]Physics Department, Western Illinois University, 1 University Circle, Macomb, IL 61455, USA
[4]National Radio Astronomy Observatory, 520 Edgemont Road, Charlottesville, VA 22903, USA
[5]Department of Astronomy and Cornell Center for Astrophysics and Planetary Science, Cornell University, Ithaca, NY 14853, USA
[6]Department of Astronomy, University of Wisconsin-Madison, Madison, WI 53706-1507, USA
[7]Department of Physics, University of Central Florida, Orlando, FL 32816, USA
[8]Green Bank Observatory, P.O.Box 2, Green Bank, WV 24944, USA
[9]Jet Propulsion Laboratory, California Institute of Technology, 4800 Oak Grove Drive, Pasadena, CA, 91109, USA
[10]Department of Astronomy, University of California, Berkeley, CA 94720-3411, USA
[11]Brigham Young University, Provo, UT 84602, USA
[12]Radio Astronomy Centre, NCRA-TIFR, Udagamandalam, India
[13]USRA, SOFIA, NASA Ames Research Center, Moffett Field, CA 94035, USA
[14]Florida Space Institute, UCF, Orlando, FL, USA
[15]Lunar and Planetary Institute, 3600 Bay Area Boulevard, Houston, TX 77058
[16]Department of Astronomy and Astrophysics, Pennsylvania State University, 525 Davey Laboratory, University Park, PA, 16802, USA